\documentclass[11pt,a4paper]{article}
\pdfoutput=1
\usepackage{jinstpub}
\usepackage[english]{babel}
\usepackage{upgreek}
\usepackage{xspace}
\usepackage{graphicx}
\usepackage[capitalize]{cleveref}
\usepackage{enumitem}
\usepackage{subcaption}
\usepackage[modulo]{lineno}
\usepackage{subcaption}
\usepackage[modulo]{lineno}
\usepackage{booktabs}
\usepackage[dvipsnames]{xcolor}
\usepackage{multirow}
\usepackage{url}
\newcommand{\gf}{\textsc{Geant4}\xspace}

\def\Offline{\mbox{$\overline{\textrm%
{Off}}$\-\hspace{.05em}\protect\raisebox{.4ex}%
{$\protect\underline{\textrm{line}}$}}\xspace}
\renewcommand{\arraystretch}{1.15}

\title{
Development and validation of the signal simulation for the underground muon detector of the Pierre Auger Observatory
}

\author[*,a,b]{A. M. Botti\note[*]{Corresponding author.},}
\author[a]{F. Sánchez,}
\author[c]{M. Roth,}
\author[a]{A. Etchegoyen}

\affiliation[a]{Instituto de Tecnolog\'\i{}as en Detecci\'on y Astropart\'\i{}culas (CNEA, CONICET, UNSAM), Buenos Aires, Argentina}
\affiliation[b]{Department of Physics, FCEyN, University of Buenos Aires and IFIBA, CONICET, Buenos Aires, Argentina}
\affiliation[c]{Karlsruhe Institute of Technology (KIT), Institute for Astroparticle Physics, Karlsruhe, Germany}

\emailAdd{ana.botti@iteda.cnea.gov.ar}

\abstract{The underground muon detector of the Pierre Auger Observatory is aimed at attaining direct measurements of the muonic component of extensive air showers produced by cosmic rays with energy from $10^{16.5}$\,eV up to the region of the ankle (around $10^{18.7}$\,eV). It consists of two nested triangular grids of underground scintillators with 433\,m and 750\,m spacings, and a total of 71 positions, each with 192 scintillator strips (30\,m$^2$) deployed 2.3\,m underground. The light produced by impinging muons in the scintillators is propagated with optical fibers towards an array of silicon photomultipliers. In this work, we present the development, validation, and performance of an end-to-end tool for simulating the response of the underground muon detector to single-muon signals, which constitutes the basis for further simulations of the whole array. Laboratory data and simulation outcomes are found consistent, showing that with the underground muon detector we can measure single muons, with an efficiency of 99\%, up to about 1050 particles arriving at exactly the same time in 30\,m$^2$ of scintillator.}

\keywords{Simulation methods and programs; Performance of High Energy Physics Detectors;
Photon detectors for UV, visible and IR photons (solid-state) (PIN diodes, APDs, Si-PMTs, G-APDs, CCDs, EBCCDs, EMCCDs, CMOS imagers, etc); Front-end electronics for detector readout}

\arxivnumber{2104.13253} 

\dedicated{\flushleft\textnormal{\textsc{%
Published in JINST as \href{https://doi.org/10.1088/1748-0221/16/07/p07059}{DOI:10.1088/1748-0221/16/07/p07059}
}}}

\begin{document}
\maketitle
\flushbottom

\section{Plastic scintillators at the Pierre Auger Observatory}
\label{sec::intro}

Cosmic rays with energies above $10^{15}$\,eV can only be indirectly detected by measuring the extensive air showers produced after they interact with nucleons of the Earth's atmosphere. The multiplicative process following the first interaction is approximately explained by the Heitler-Matthews model of hadronic cascades~\cite{Matthews:2005sd}: secondary particles are produced by the decay or interaction of products from the first interaction. In this way, we can identify two distinguished components in the shower, namely the hadronic and the electromagnetic components. The latter is mainly composed of electrons, positrons, and photons, the former by predominantly pions and kaons; the decay of kaons and electrical charged pions constitutes the main contribution to the air shower muons. Even though the Heitler-Matthews model is a simplified picture of the real process involved in the extensive air shower development, it is accurate enough to replicate three major features: (i) the atmospheric depth of the maximum development of the cascade is linearly dependent on the logarithm of the energy, $E$, of the primary cosmic ray, (ii) the number of muons in the shower, $N_\upmu$, grows as $(E/\xi_\mathrm{c})^\beta$ and (iii) the mass number $A$ of the primary particle scales as $N_\upmu\propto\ A^{\beta-1} (E/\xi_\mathrm{c})^\beta$ where $\mathrm{\xi_\mathrm{c}}$ is the critical energy at which charged pions decay into muons\footnote{Kaons decay into charged pions, which in turn, decay into muons.} rather than interacting with nucleons, and $\mathrm{\beta\approx 0.9}$ is an effective parameter of the model~\cite{AlvarezMuniz:2002ne}.

In the Heitler-Matthews model, the muon content of an extensive air shower is directly related to the mass of the primary particle, thus, direct measurements of these muons with ground-based arrays have become increasingly relevant to unlock the riddles concerning cosmic rays with energy above $10^{15}$\,eV, the so-called high- and ultra-high-energy ($\geq 10^{18}$\,eV) cosmic rays. Both the consistency of hadronic interaction models for these energies tuned with the Large Hadron Collider data~\cite{MuonDeficit}, as well as the astrophysical scenarios still compatible with observations~\cite{AugerSpectrum}, can only be explored knowing in more detail the mass composition of the cosmic ray flux. Precise and direct measurements of muons at the ground are, undoubtedly, of utmost importance to forward our understanding of the physics of high- and ultra-high-energy cosmic rays.

The implementation of plastic scintillator arrays is a common approach to measure the lateral profile of air shower particles. Their versatility allows the deployment of these detectors either on the ground or buried underground. On-ground, they are optimal to measure the most abundant electromagnetic component of the cascades, whereas underground, they are ideal to measure directly the shower muons as electrons, positrons, and photons are easily shielded by the soil overburden. To improve the sensitivity in large-area scintillator detectors, optical fibers are typically used to collect the light produced in the plastic and conduct it to a photo-sensor that might be meters apart from the impinging position. In this way, it is possible to use large scintillator surfaces achieving better control of fluctuations in the measurements of the particle density in air showers.

At the Pierre Auger Observatory~\cite{PierreAugerObservatory}, cosmic rays with energies above $10^{16.5}$\,eV are measured with a hybrid detection technique. Arrays of water-Cherenkov detectors covering different areas and different spacing are used to measure the lateral spread of shower particles at the ground. These arrays, commonly referred to as the surface detector, are deployed in triangular grids with spacings of 1500\,m (SD-1500), 750\,m (SD-750) and 433\,m (SD-433) for energy thresholds of about $10^{18.5}$\,eV, $10^{17.5}$\,eV, and $10^{16.5}$\,eV respectively. Additionally to the surface detector, which has a 100\% duty cycle, the Pierre Auger Observatory is also equipped with a fluorescence detector consisting of 27 telescopes to measure the fluorescence light produced with the longitudinal development of the air showers in the atmosphere. It allows for a measurement of the calorimetric energy of the primary particle and it is optimal to determine the depth of the maximum development of the extensive air shower on an event-by-event basis. Nevertheless, as it can be operated only during moonless nights, it has a reduced duty cycle of around 15\%.

\begin{figure} [t]
\centering
\includegraphics[trim={0.0cm 0.0cm 0.0cm 0.0cm},clip,width=1.0\textwidth]{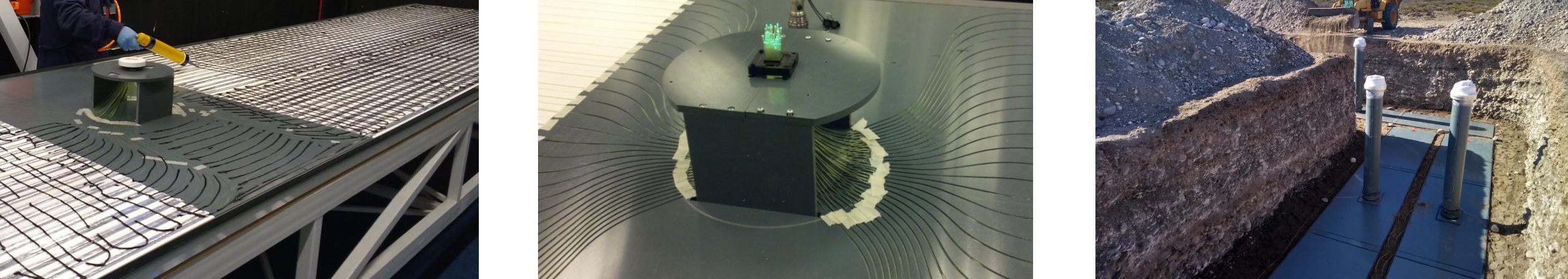}
\caption{UMD module under construction (left and middle) and a UMD station after deployment and before being overburden with soil (right).
\label{fig:deployment}}
\end{figure}

As part of its upgrade, dubbed AugerPrime~\cite{AugerPrime}, arrays of plastic scintillator detectors are being deployed at the Pierre Auger Observatory on top of each water-Cherenkov detector of the SD-1500 to measure air shower muons and electromagnetic particles for the highest-energy cosmic rays. The surface scintillator detector (SSD) will provide a complimentary response from which the different components of the signals, roughly distinguished by their electromagnetic or hadronic origin, can be deconvolved. In addition, direct access to the muonic content of air showers at energies up to the ankle region (around $10^{18.7}$\,eV conserving high statistics)\footnote{This upper limit is a consequence of the flux and the area covered by the SD-750 and SD-433.}, will be attained with arrays of plastic scintillator detectors deployed at 2.3\,m underground and at each position of the SD-750 and SD-433~\cite{MuonsWithAMIGA}. The underground muon detector (UMD) will provide clean and precise measurement of the muons in the extensive air shower improving the cosmic ray mass identification in the second-knee and ankle regions of the energy spectrum~\cite{spectrum}. Each SSD station consists of 48 plastic-scintillator strips of (160\,$\times$\,5\,$\times$\,1)\,cm with wavelength-shifting optical fibers deployed in an aluminum container constituting two modules with a total active area of 3.8\,m$^2$~\cite{SSD}. The read-out electronics consist mainly of a photomultiplier tube with low- and high-gain outputs. On the other hand, each UMD station consists of three 10\,m$^2$ modules segmented into 64 plastic-scintillator strips of (400\,$\times$\,4\,$\times$\,1)\,cm also with wavelength-shifting optical fibers that conduct the light from the scintillators to an array of 64 silicon photomultipliers (SiPMs)~\cite{AMIGAPrototype, AMIGASIPM}. Pictures of a UMD module construction (left and middle) and of a station deployment (right) are presented in Fig.~\ref{fig:deployment}. We paint in black the optical-fiber end that is not coupled to the SiPM to avoid photon reflections and, in turn, the arrival of delayed photons to the SiPM.

In this paper, we present a detailed description of the simulation sequence of one scintillator strip with the optoelectronics system of the UMD based on a thorough characterization of signals measured in the laboratory with individual strips. The setup built in the laboratory includes a complete replica of the UMD acquisition chain under controlled conditions which has allowed us to both fine-tune the parameters used in the simulation and to validate their outcome. The structure of this article is as follows: section~\ref{sec:setup} is an introduction of the UMD while section~\ref{sec:simulations} is devoted to the explanation of the simulation rationale; in section~\ref{sec:validation} the characterization of single-muon signals and the validation of the simulation in terms of the efficiency and dynamic range of the detector are described; finally, we summarize and conclude in section~\ref{sec:conclusion}.

\section{The underground muon detector and its simulation}
\label{sec:setup}

When a muon impinges on a scintillator strip, photons are generated and propagated through the optical fiber to a SiPM\footnote{In the UMD, the photo-detectors are 1584-cell SiPM, Hamamatsu S13361-2050.} producing a single-muon signal. The SiPM output is then processed with two different read-outs, corresponding to each of the acquisition modes of the UMD~\cite{UMDFrontEnd, UMDCalibration, UMDComms}: the {\it binary} and {\it ADC} modes implemented to probe muon densities ranging from one up to tens of muons per $\mathrm{m^2}$. The binary acquisition is designed to measure low particle densities, by attaining a high efficiency for single particles and reducing the signal fluctuations in this range. It handles independently the 64-SiPM signals through a pre-amplifier, fast-shaper, and a discriminator, built within each channel of two 32-channel Application-Specific Integrated Circuits (ASIC). The maximum number of simultaneous particles that can be detected with this mode is limited by its own segmentation as it can only detect one muon per strip at the same time. This limit is extended with the ADC mode, in which the 64 SiPM signals are summed and afterward amplified.

\begin{figure}[t]
	\centering
	\begin{subfigure}{1.00\textwidth}
		\centering
		\includegraphics[trim={0.0cm 0.0cm 0.0cm 0.0cm},clip,width=1.0\linewidth]{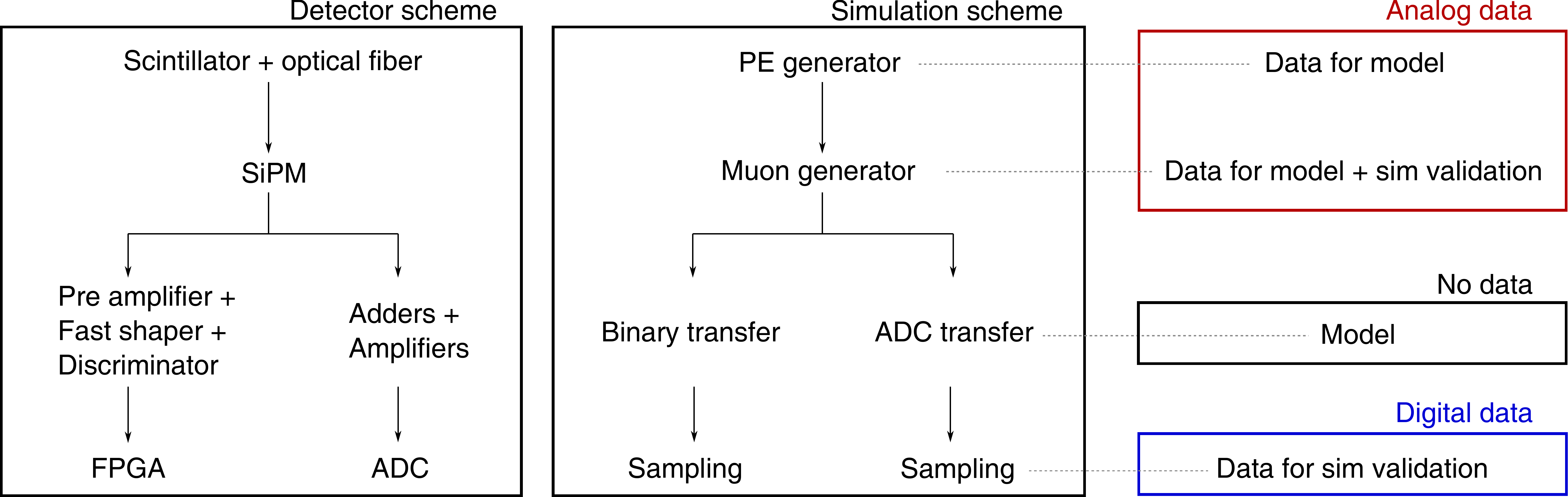}
	\end{subfigure}
	\caption{Schematics of the detector working chain (left), simulation chain of the single-muon signal (middle), and layout of the laboratory data used to develop and validate the simulation (right).}
	\label{fig:electronics}
\end{figure}

At the output of a single UMD module, we obtain 64 traces for the binary mode and two for the ADC mode. The memory depth of the UMD electronics stores 6.4\,$\upmu$s; each trace of the binary mode has 2048 bits where a ``1'' or a ``0'' is output if the fast-shaper signal is above or below the discriminator threshold. For the ADC mode, on the other hand, the outcomes are waveforms of 1024 samples. The binary and ADC signals are sampled in two different ways: (a) the discriminator signals of the binary mode with a Field-Programmable Gate Array (FPGA) at a speed of 320\,MHz (3.125\,ns sample time), and (b) the ADC mode with two Analog-to-Digital Converters (ADCs) at a speed of 160\,MHz (6.25\,ns sample time). A schematic flow of the working chain corresponding to one scintillator strip is displayed in the left panel of Fig.~\ref{fig:electronics}.

To translate the signal of the UMD modules into a particle density, both binary and ADC traces need to be converted to a number of impinging muons. This implies interpreting the number of photons collected by SiPMs in the form of photo-equivalents\footnote{We refer to as "photo-equivalents" to the number of triggered cells in the SiPM.} (PE) processed by the electronics after a charged particle passes through the scintillator and photons propagate through the optical fiber. The shape of the final traces depends on the arrival time profile of the photons, and in turn on the timing of the excitation and de-excitation processes of both scintillator and optical fiber molecules. In the ADC acquisition mode, we use the signal charge to infer the number of impinging muons, and therefore the PE timing is not relevant in the estimation of the muon density. However, when we count and store signals above a given amplitude, as in the binary mode, the timing of PEs plays a key role, since the amplitude of the signal is determined by the superposition of the PE pulses according to their arrival time profile.

PE pulses are not only the outcome of incoming light; they are also produced by thermal fluctuations within the SiPM cells when electrons spontaneously jump to the silicon conduction band and trigger an avalanche identical to that generated by an impinging photon. Furthermore, when electrons in a triggered cell (disregarding if it was by incident radiation or thermal noise) are recombined, the photons produced may escape towards a neighboring cell inducing a second and simultaneous avalanche. This effect, known as optical cross-talk, may produce pulses of more than one PE, even though only one cell was triggered either by impinging radiation or thermal noise and is accounted for in the simulation.

Each scintillator strip of the UMD is simulated following the steps illustrated in the middle panel of Fig.~\ref{fig:electronics}. The sequence is based on the parameters accessible (and therefore useful to tune and validate the simulation) with laboratory measurements: analog 1\,PE pulses produced by thermal fluctuations in the SiPM, analog single-muon pulses, the binary traces, and the ADC waveforms obtained with the electronics. These last three, produced at fixed distances over the scintillator strips and, thus, giving information on the effect of the light attenuation along with the optical fiber. Taking this into account, the main ingredients to simulate the sequence for single particle consists of:

\begin{enumerate}
\item a PE pulse generator developed to simulate the signals of individual photons arriving at the SiPM,
\item a single-muon generator that sums several PE signals according to a given number of photons arriving at the SiPM with a time profile that follows the physical processes involved in the photon generation and propagation,
\item a transfer model corresponding to the read-out electronics of the binary and ADC modes,
\item a sampler for the final digitization and discretization of the analog signals for both acquisition modes.
\end{enumerate}

\begin{figure} [t]
\centering
\includegraphics[trim={0.0cm 0.0cm 0.0cm 0.0cm},clip,width=0.9\textwidth]{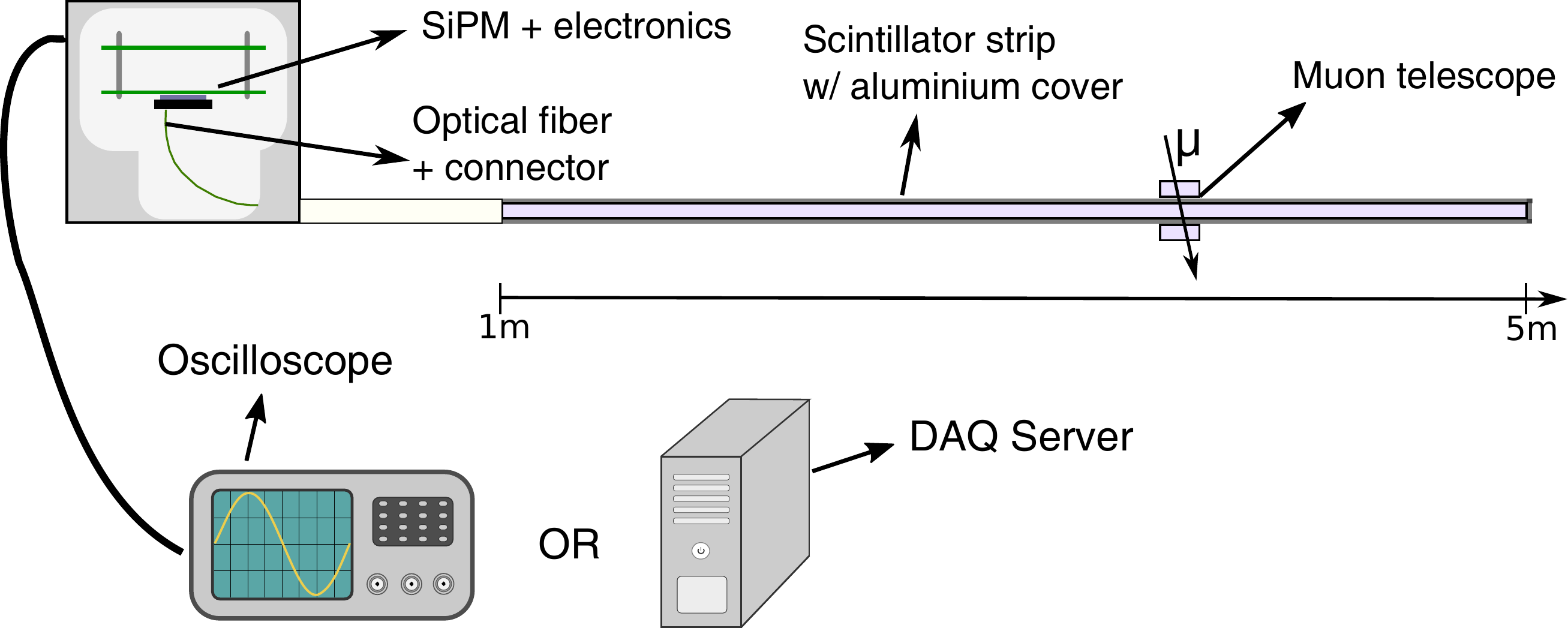}
\caption{Schematics of the laboratory setup used to extract analog and digital single-muon signals.
\label{fig:pipaDesign}}
\end{figure}

To acquire the laboratory data used to develop, tune, and validate the simulation, we used the setup illustrated in Fig.~\ref{fig:pipaDesign}: one scintillator strip with its optical fiber coupled to a SiPM, all components identical to those used in UMD modules~\cite{AMIGAPrototype}. To trigger the acquisition, a muon telescope consisting of two (4\,$\times$\,4\,$\times$\,1)\,cm scintillators, matching the width of one scintillator strip, was used. When a muon impinges on the segments of the telescope, a trigger signal for acquisition is produced. Moving the telescope over the length of the strip allowed us to obtain muon signals as a function of the optical-fiber length between the particle impact position and the SiPM and, therefore, to retrieve the information of the light attenuation in the fiber. The SiPM output was processed with two different read-out systems: a customized electronics plus an oscilloscope to obtain analog signals, and standard UMD electronics to obtain digitized signals.

\begin{figure}[t]
	\centering
	\begin{subfigure}{0.8\textwidth}
		\centering
		\includegraphics[trim={0.0cm 0.0cm 0.0cm 0.0cm},clip,width=1.0\linewidth]{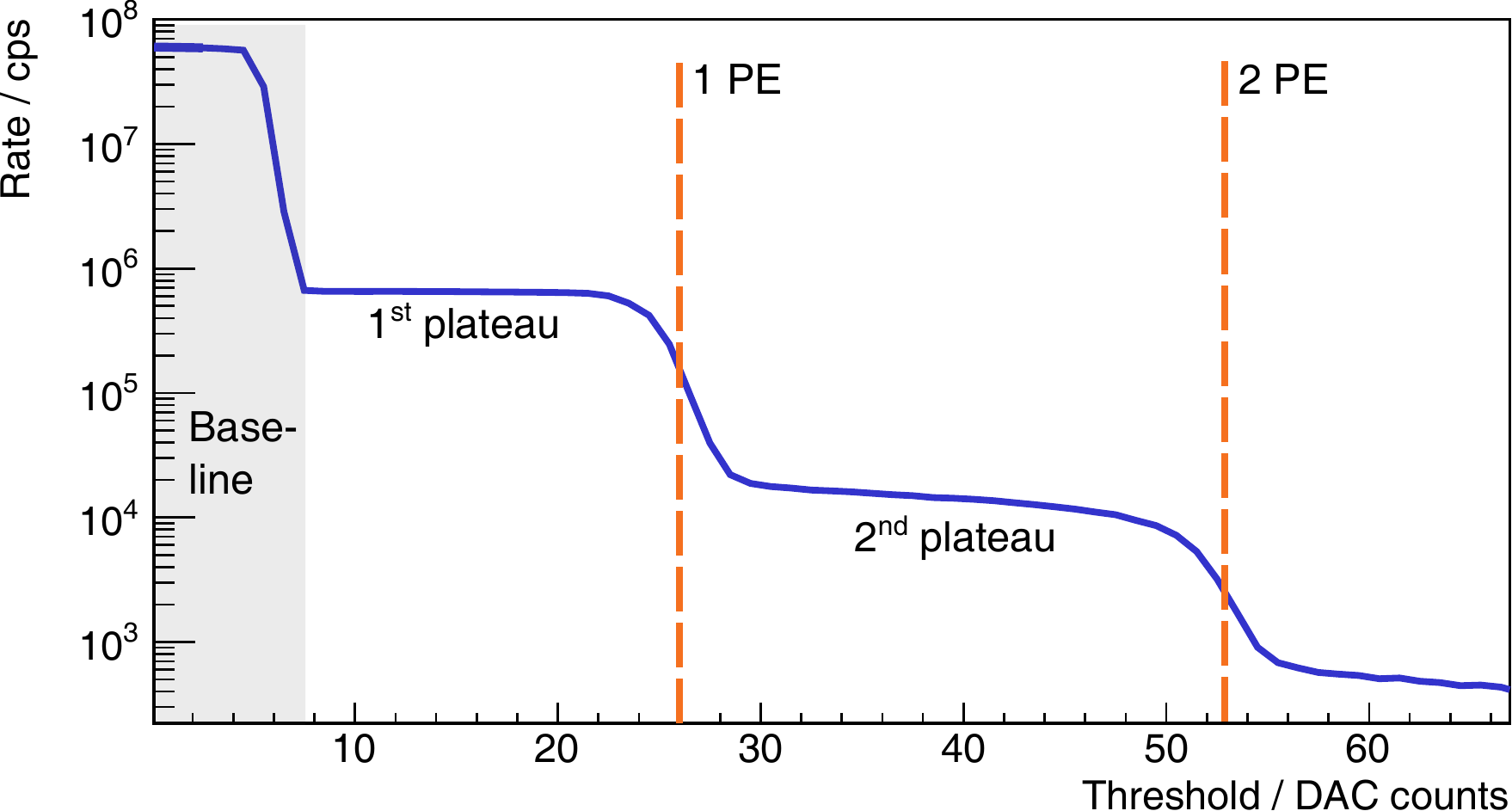}
	\end{subfigure}
	\caption{Dark-count rate in counts per second (cps) as a function of the discriminator threshold acquired with an over-voltage of 3.5\,V. 
	Orange-dashed lines indicate the position of the 1 and 2\,PE peaks. The gray area denotes the region where the baseline is found.}
	\label{fig:calibrationPEGen}
\end{figure}

The SiPM reverse-bias voltage, $V_\mathrm{{bias}}$, in the laboratory setups was set following the procedure explained in~\cite{UMDCalibration}, so the configuration used in the Pierre Auger Observatory is properly reproduced. The procedure consists of determining the SiPM break-down voltage, $V_\mathrm{{br}}$, by means of the binary channel and then setting the reverse-bias voltage to fix the same over-voltage ($V_\mathrm{{ov}}$\,=\,3.5V) in all SiPMs: $V_\mathrm{{bias}} = V_\mathrm{{br}} + V_\mathrm{{ov}}$. The procedure relies on obtaining the 1\,PE signal amplitude as a function of the $V_\mathrm{{bias}}$ by measuring the rate of dark counts: for each $V_\mathrm{{bias}}$, we count the number of transitions in the binary trace between a ``0'' and a ``1'' for different discriminator thresholds. As a result, for each $V_\mathrm{{bias}}$, we obtain the dark-count rate as a function of the discriminator threshold as presented in Fig.~\ref{fig:calibrationPEGen}. In other words, this is the number of pulses per second at the fast-shaper output crossing the discriminator threshold as a function of this threshold. For each integer number of PE, the signal amplitude is well separated from its consecutive PE, therefore, when the discriminator threshold moves between two PE peaks, the rate of dark counts varies slowly and a plateau is obtained; when the threshold moves close to the mean signal amplitude of an integer number of PEs, the rate varies rapidly and a drop in the rate is obtained. As a consequence, the PE mean amplitude can be easily identified in the transition between two consecutive plateaus. This method, based on measuring the SiPM thermal noise, was developed to set the operating point of the UMD modules to guarantee a uniform response in terms of noise and gain among the whole array. We indicate the positions of the first and second plateaus along with the mean amplitude of the one- and two-PE signals (indicated with dashed-orange lines). 

As we mentioned in Fig.~\ref{fig:electronics}, analog data obtained in the laboratory was used to develop both the single-PE and single-muon pulse generator: we obtained two thousand dark-count signals with 1\,PE amplitude, and 16000 single-muon signals using the muon telescope at eight positions on the scintillator strip with steps of 0.5\,m. This latter dataset was also used to validate the muon generator by characterizing the amplitude, charge, and width of the single-muon analog signals. In this same way, digital data was obtained to validate the output of the whole optoelectronic simulation chain.

\section{Simulations}
\label{sec:simulations}


The simulation of the UMD scintillator strip was not developed using models based on first principles or a microscopic analytical assumption of the detector physics, but through a phenomenological approach that describes the macroscopic behavior of the detector components tuned to laboratory data. This methodology allowed for a simulation that accurately represents the signal features of the detector minimizing the computing costs. The steps corresponding to the photon production in the scintillator, propagation through the optical fiber, and SiPM response were developed using analog measurements. In these steps, model parameters were extracted by fitting laboratory data, and used as the input of a pulse generator. Then, the response of the read-out electronics and subsequent sampling were simulated using models for the transfer functions tuned to digital data also acquired in the laboratory. As a result, each step of the simulation sequence was validated by means of the laboratory data to guarantee an accurate and realistic representation of the main features of the UMD traces and, therefore, of the full detector response. We present in the following subsections the simulation of the analog and digital signals of UMD.

\subsection{Analog simulation}
\label{subsec:simDR}

\begin{figure}[t]
       \hspace{-0.2cm}
	\centering
	\begin{subfigure}{.49\textwidth}
		\centering
		\includegraphics[trim={0.0cm 0.0cm 0.0cm 0.0cm},clip,width=1.0\linewidth]{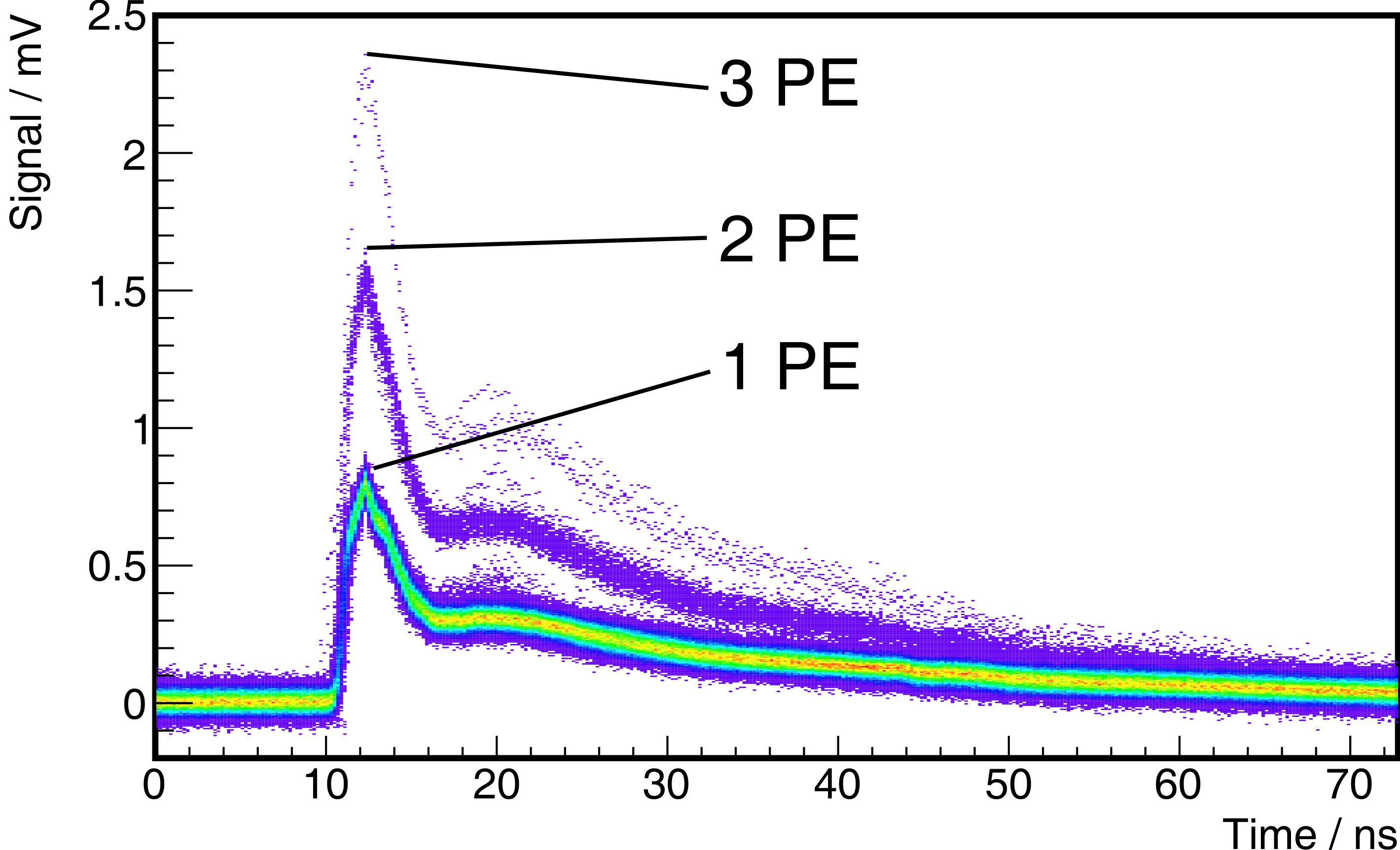}
	\end{subfigure}
	\hspace{0.2cm}
	\begin{subfigure}{0.49\textwidth}
		\centering
		\includegraphics[trim={0.0cm 0.0cm 0.0cm 0.0cm},clip,width=1.0\linewidth]{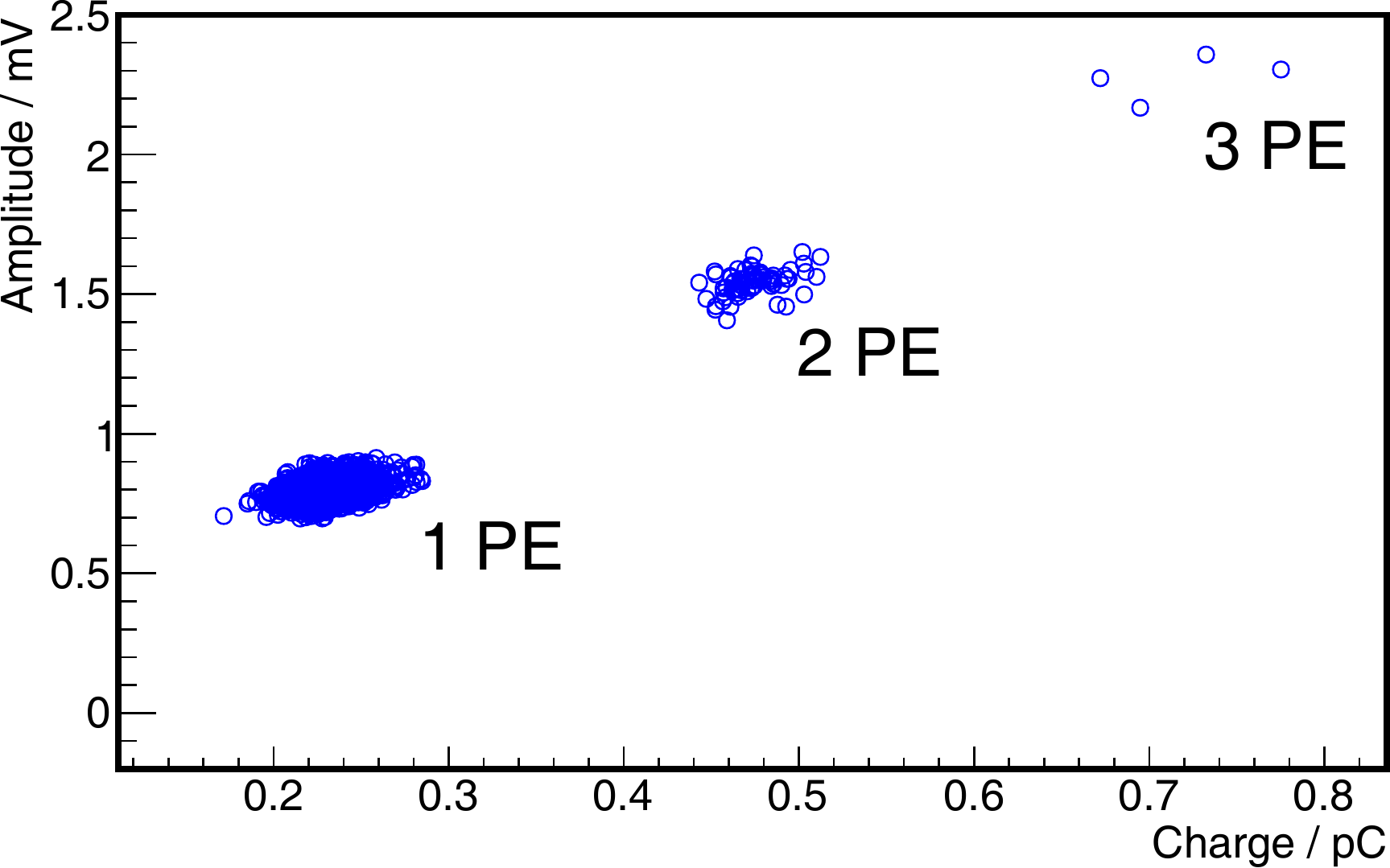}
	\end{subfigure}
	\caption{(Left) overlapped dark-count pulses. We can clearly identify the different PE peaks produced by crosstalk between the SiPM inner-cells. (Right) signal amplitude as a function of its charge (signal integral). The three populations correspond to the one-, two- and three-PE signals.}
	\label{fig:sipmSignalsRaw}
\end{figure}

When a muon impinges on a scintillator strip, scintillation photons are produced in the plastic and subsequently propagated through the optical fiber to the SiPM resulting in a signal; this is the addition of the pulses produced by each individual photon. For this reason, the first practical step towards the simulation of the UMD signal, as displayed in Fig.~\ref{fig:electronics}, is the development of a PE generator to reproduce the SiPM analog pulses for individual photons. To this aim, we used PE pulses obtained in the laboratory and presented in the left panel of Fig.~\ref{fig:sipmSignalsRaw}. Since the signal produced by dark counts is indistinguishable from that by impinging photons, this data set was built using thermal noise. In the figure, we display an overlap of all the data, where the effect of the optical cross-talk is apparent: the one-, two-, and three-PE signals are clearly visible. To better illustrate this, we present in the right panel of Fig.~\ref{fig:sipmSignalsRaw} the maximum amplitude of signals as a function of their charge (corresponding to the signal integral), where each population corresponds to a plateau transition previously presented in Fig.~\ref{fig:calibrationPEGen}, and in turn, to each integer number of PEs. It is also worth mentioning that for the 1\,PE population, the amplitude and charge means are (0.798\,$\pm$\,0.001)\,mV and (0.2313\,$\pm$\,0.0001)\,pC with a standard deviation of (0.0327\,$\pm$\,0.0004)\,mV, and (0.0117\,$\pm$\,0.0002)\,pC. The fluctuations of about 5\% contain all constructive differences within each of the 1584 SiPM cells and denote the uniformity in the response between SiPMs.

The PE generator was built using a phenomenological model for the signal of individual photons. To fit the dark-count pulses displayed in Fig.\ref{fig:sipmSignalsRaw}, we used a function with one exponential rise time and three exponential fall times~\cite{maranoElectric}

\begin{equation}
\operatorname{Signal}\left(t\right) = A_1 \left(1-e^{\frac{-t}{\tau_\text{r}}}\right) \left(A_2\,e^{\frac{-t}{\tau_{1}}}+A_3\,e^{\frac{-t}{\tau_{2}}}+e^{\frac{-t}{\tau_{3}}}\right),
\label{eq:electricmodel}
\end{equation}

\noindent where $A_1$, $A_2$, $A_3$ are the coefficients that determine the signal amplitude, $\tau_\text{r}$ is the rise time, and $\tau_{1}$, $\tau_{2}$, and $\tau_{3}$ are the three fall times. The first corresponds to the quenching timing; the second, and slow, decay corresponds to the recovery time of the SiPM cells. The third corresponds to a fast decay which is produced by the coupling of parasitic capacitors to the load impedance of the read-out system~\cite{sipms101, maranoElectric}. The parameters of the model are estimated by fitting the measured 1\,PE pulses, for which signals that not exclusively contained single-PEs, were removed by dismissing all events outside the first population in the right panel of Fig.~\ref{fig:sipmSignalsRaw}.

To properly model the signal fluctuations, each measured PE pulse is fitted with Equation~\eqref{eq:electricmodel}, and a histogram is filled for each parameter to extract its corresponding mean and standard deviation. Results are summarized in Table~\ref{table:params}. The fluctuations of the signal baseline were estimated using the first 100 samples of the individual PE signals, obtaining a mean fluctuation of (26.90 $\pm$ 0.04)\,$\upmu$V with a standard deviation of (1.78 $\pm$ 0.03)\,$\upmu$V, and included as a parameter of the simulation.
The PE generator is, therefore, a set of random generators with normal distributions whose means and standard deviations are the values presented in Table~\ref{table:params}.

\begin{table}[t]
\centering
\renewcommand{\arraystretch}{1.25}
 \begin{tabular}{c c c}
\toprule
 Parameter & Mean & Std. Dev.\\
\midrule
 $A_1$ / mV & 0.29 $\pm$ 0.01 & (1.37 $\pm$ 0.03) $10^{-2}$\\
 $A_2$ & 23.22 $\pm$ 0.07 & 2.94 $\pm$ 0.06 \\
 $A_3$ & 1.609 $\pm$ 0.001 & 0.054 $\pm$ 0.001\\
 $\mathrm{\tau_r}$ / ns & 3.82 $\pm$ 0.01 & 0.62 $\pm$ 0.01 \\
 $\mathrm{\tau_{1}}$ / ns & 1.187 $\pm$ 0.002 & 0.070 $\pm$ 0.001\\
 $\mathrm{\tau_{2}}$ / ns & 23.44 $\pm$ 0.04 & 1.95 $\pm$ 0.04\\
 $\mathrm{\tau_{3}}$ / ns & 0.221 $\pm$ 0.001 & (3.24 $\pm$ 0.04) $10^{-2}$\\
\bottomrule
 \end{tabular}
 \caption{Mean and standard deviation for the parameters in Equation~\eqref{eq:electricmodel} obtained by fitting two thousand signals of 1\,PE.}
 \label{table:params}
\end{table}

\begin{figure}[t]
	\centering
	\begin{subfigure}{0.8\textwidth}
		\centering
		\includegraphics[trim={0.0cm 0.0cm 0.0cm 0.0cm},clip,width=1.0\linewidth]{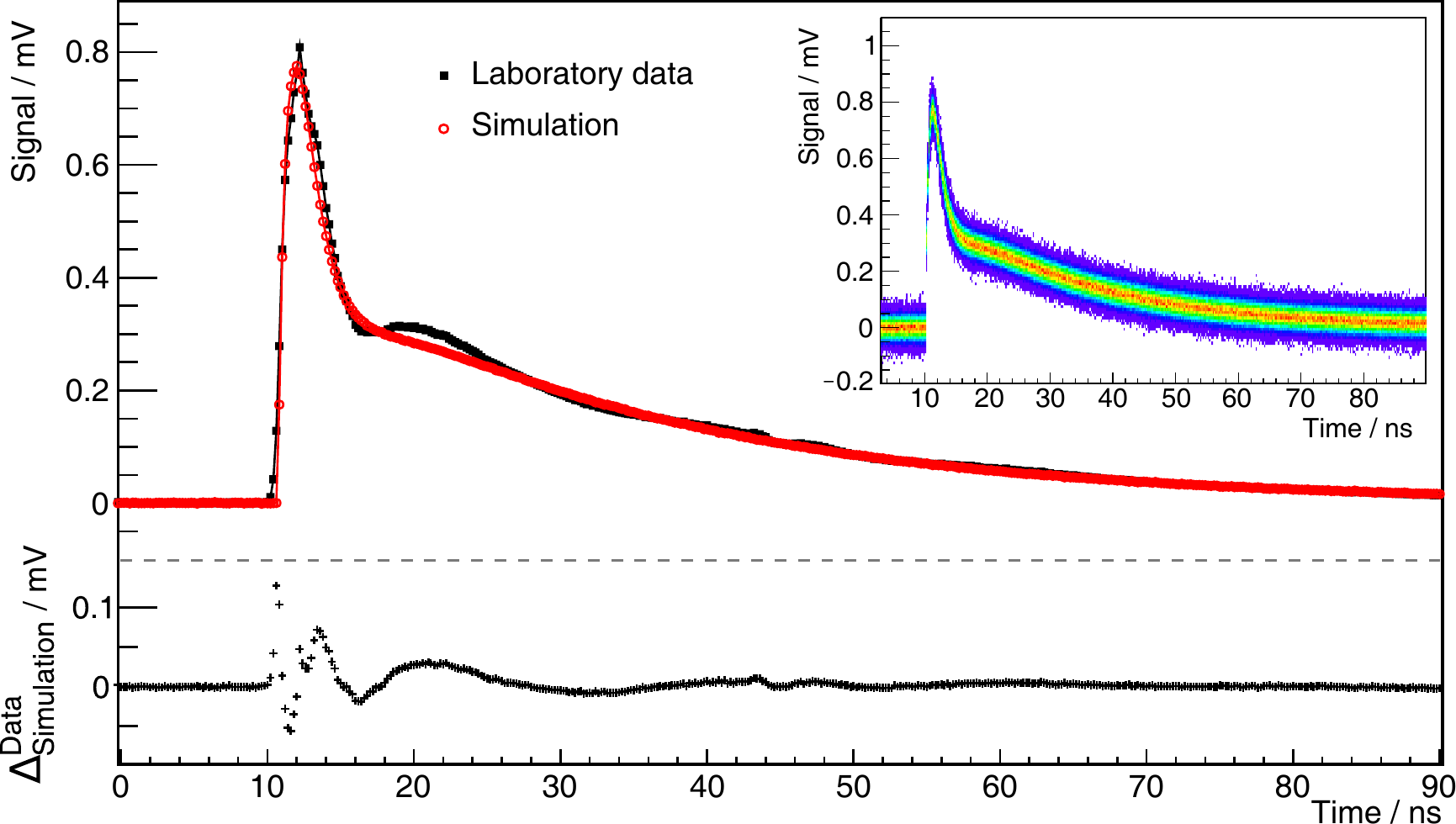}
	\end{subfigure}
	\caption{Average single-PE signal of two thousand pulses for both laboratory data and simulation. At the bottom, we present the difference between the averages at each sample time. In the inset, we display the two thousand simulated pulses obtain with the PE-generator.}
	\label{fig:residuals}
\end{figure}

To test the accuracy of the PE generator, we created a data set with two thousand simulated single PEs and compared it with the measured dark counts used to develop the generator. The mean signal for each data set was obtained by averaging all the pulses in the data set, and the results are presented in Fig.~\ref{fig:residuals}, for both laboratory data (black squares) and simulation (red circles). The difference between the two waveforms is displayed in the bottom of Fig.~\ref{fig:residuals}; it varies between $-$5\% and 10\% of the measured signal, having the maximum differences when the signal changes abruptly, i.e.\,at the signal rise and first fall, and the following bump. However, it is worth noting that the difference oscillates around 0, and the average is quite negligible: the difference in the integrated signal charge is less than 1\% and less than 3\% when looking at the maximum amplitude, being these the most relevant parameters of the UMD performance. Each of the simulated PE pulses, presented in the inset of Fig.~\ref{fig:residuals}, was fitted in the same way as in the laboratory data, and the parameter mean and standard deviation were computed. We then obtained a maximum difference of 9\% between laboratory measurements and simulation, with differences mostly below 5\%, which proves the accuracy of the proposed algorithm.

After devising a generator that accurately simulates the single-PE pulses, we can reproduce the muon signal by adding several PEs. When a muon impinges on the detector, photons are produced in the plastic scintillator and then absorbed and re-emitted by the optical fiber; all three processes have their own characteristic times and spectral distributions. The photons re-emitted in the fiber are propagated towards the SiPM and finally detected with an efficiency that depends on their wavelength and is about 40\% at 492\,nm. Due to the emission delays, the incoming photons spread in time, producing a signal with a structure for which the number of PEs in amplitude is not equivalent to the number of PEs in charge, since not all photons add to the maximum signal amplitude. Furthermore, the total number of PEs measured by the SiPM depends on the position where the muon arrived at the scintillator strip due to the optical-fiber attenuation, which follows a double-exponential decay law~\cite{attenuation}

\begin{equation}
\overline{n_\mathrm{PE}} = n_\mathrm{PE}(0)\left( a\,\mathrm{e}^{-\frac{x}{\lambda_1}}+(1-a)\,\mathrm{e}^{-\frac{x}{\lambda_2}}\right),
\label{eq:attenuation2}
\end{equation}

\noindent where $\lambda_1$, $\lambda_2$ are the attenuation lengths, \textit{a} is a proportionality constant, and $n_\mathrm{PE}(0)$ corresponds to the mean number of PE that would be detected if a muon impinged the scintillator at 0\,m from the SiPM.
This model, with a short and long attenuation length, describes well the physical processes involved in the scintillator and optical fiber system.  Nevertheless, since measurements were taken only at distances greater than 1\,m in this work, the contribution of the short attenuation length, which is a few centimeters, is negligible. Therefore, we fixed $\lambda_2=3.5$\,cm and considered $n_\mathrm{PE}(0)$, $\lambda_1$ and $a$ as the
only free parameters of  Equation~\eqref{eq:attenuation2}.
We display the number of PEs by (i) dividing the signal maximum amplitude by the maximum amplitude of 1\,PE (blue squares), and (ii) by dividing the signal charge by the mean charge of 1\,PE (orange circles). 
Fitting results are displayed on tables in the left panel.
As previously mentioned, given the time profile of the arriving photons, not all PEs add to the maximum signal amplitude, even though they add to the total signal charge; this explains the difference between the two curves. It is worth noting that the binary mode, as it implements an amplitude threshold, is sensitive to the amplitude attenuation, while the ADC mode is sensitive to that in charge. With the parametrization in Equation~\eqref{eq:attenuation2}, we obtain the mean and variance of the number of PEs for vertical muons at each position on the scintillator strip without requiring a detailed simulation of the energy deposit in the plastic or of the subsequent photo-production and propagation through the optical fiber. Nevertheless, it is important to stress that, when merging the code into the Auger official simulation
framework (\Offline)~\cite{offline},  non-vertical muons and the energy deposit fluctuations are simulated using \gf~\cite{geant4}. In contrast, the photon production in the scintillator and optical fiber propagation are parametrized. This approach significantly simplifies the complexity of the simulation algorithm and drastically reduces the computing time. Furthermore, the parametrization of the number of PEs already includes the effects of the SiPM photo-detection efficiency and cross-talk, which also aids in improving the simulation performance.

\begin{figure}[t]
	\begin{subfigure}{0.65\textwidth}
		\includegraphics[trim={0.0cm 0.0cm 0.0cm 0.0cm},clip,width=1.0\linewidth]{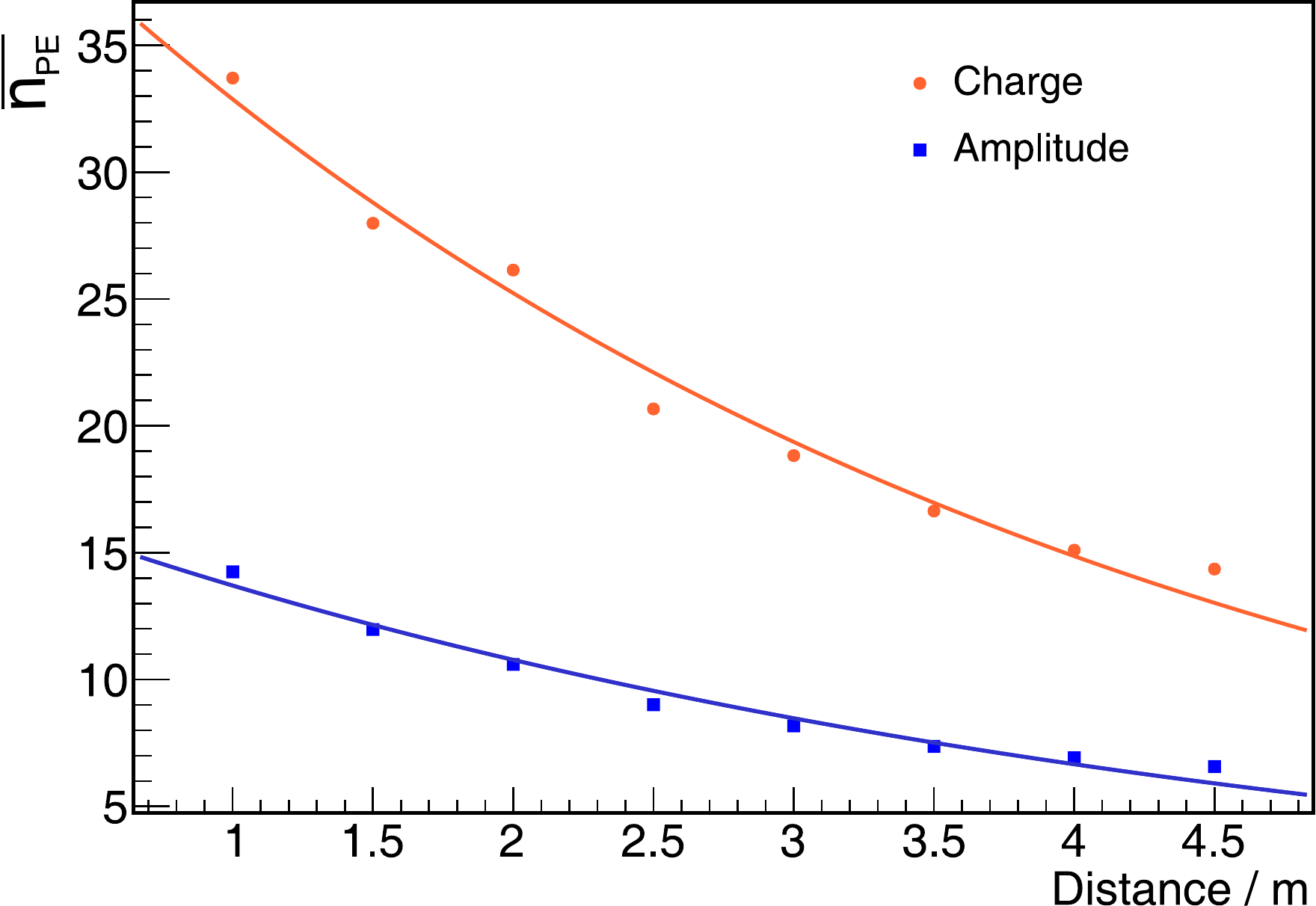}
	\end{subfigure}
	\hspace{1.0cm}
	\begin{subfigure}{.1\textwidth}
     \begin{tabular}{c c}
    \toprule
     {\color{BurntOrange} {\huge ---}} & Charge\\
    \midrule
     $n_\mathrm{PE}(0)\ /\ \text{PE}$ & 43 $\pm$ 2\\
     $a$ & 0.97 $\pm$ 0.01\\
     $\lambda_1$ / m & 3.8 $\pm$ 0.3\\
     $\lambda_2$ / m & 0.035 (fixed)\\
     \bottomrule \\
     \toprule
     {\color{blue} {\huge ---}} & Amplitude\\
     \midrule
     $n_\mathrm{PE}(0)\ /\ \text{PE}$ & 17.4 $\pm$ 0.7 \\
     $a$ & 0.95 $\pm$ 0.01 \\
     $\lambda_1$ / m & 4.2 $\pm$ 0.3 \\
     $\lambda_2$ / m & 0.035 (fixed) \\
    \bottomrule
     \end{tabular}
	\end{subfigure}
	\caption{Mean number of photo-equivalents in charge and amplitude as a function of the telescope position. Data were fitted with the double-exponential decay function in Equation~(\ref{eq:attenuation2}) with $n_\mathrm{PE}(0)$, $\lambda_1$ and $a$ as the
only free parameters; fit results are displayed in the tables on the right.}
	\label{fig:average}
\end{figure}

\begin{figure}[t]
	\centering
	\begin{subfigure}{1.00\textwidth}
		\centering
		\includegraphics[trim={0.0cm 0.0cm 0.0cm 0.0cm},clip,width=1.0\linewidth]{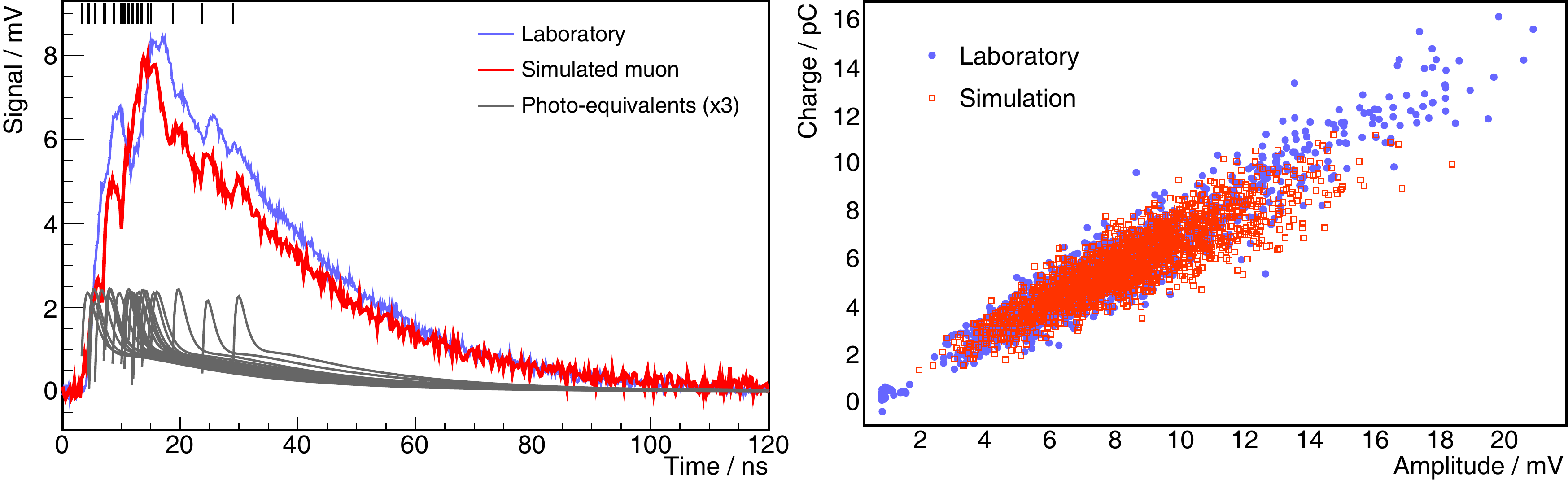}
	\end{subfigure}
	\caption{(Left) example of simulated muon signal at 2\,m of the scintillator strips (red), with its single-PE pulses re-scaled to ease the visualization (gray), along with measured signal at the same position (blue). Ticks on the top x-axis represent the start time of the PE pulses. (Right) correlation between the muon signal charge and amplitude for two thousand events of both laboratory data and simulation.}
	\label{fig:muonPulse}
\end{figure}

The single-muon generator is, therefore, a random generator following a Poissonian distribution whose parameter is the mean number of PEs
measured in a given position on the scintillator strip,
extracted from the charge attenuation parametrization presented in Fig.~\ref{fig:average}.
Having drawn the number of PEs,
we then generate for each a pulse with the PE generator, described previously.
The start time of each PE signal is determined using two random generators following exponential decays, with the characteristic decay times of the scintillator and optical fiber. By adding all the PE signals within the corresponding time profile, the single-muon signal is obtained. As an example, we present in the left panel of Fig.~\ref{fig:muonPulse} a simulated muon signal at 2\,m on the scintillator strip (red) with its individual PEs (gray) re-scaled to ease the visualization: the different start times of the PEs determine the time structure of the muon signal. In the same figure, we also present, only for illustrative purposes, a selected signal measured in the laboratory also at 2\,m (blue).

The correlation between the muon signal charge and its amplitude depends on the arriving time of the photons, which in turn, determines the signal correlation between binary and ADC channels. In this sense, it is of utmost importance to provide an accurate simulation of the correlation between amplitude and charge, as we present in the right panel of Fig.~\ref{fig:muonPulse}. The signal charge as a function of its maximum amplitude is presented for measured (blue circles) and simulated (red squares) muon signals at 2\,m on the scintillator strip. Both simulations and measurements are consistent around the means, demonstrating that the amplitude: charge correlation is directly detached from the proper simulation of the photon time profile. A slight difference of less than 5\% in the slopes is identified, along with outliers in laboratory data for high and low amplitudes that are not reproduced in the simulation. Nevertheless, these represent roughly 7\% of the data, and, as it will be presented in Section~\ref{sec:validation}, does not have an impact on the simulation accuracy. Furthermore, these differences are expected since the number of PEs in the simulation is extracted using Equation~(\ref{eq:attenuation2}), and large fluctuations in the energy deposit produced from the landau tail are not considered.


The completion of a generator for single-muon signals concludes the analog part of the simulation. The following steps, as presented in the next Subsection, consists of the electronics processing and sampling for which transfer models and digital data were used.

\subsection{Digital simulation: binary and ADC acquisition modes}
\label{subsec:simeKIT}

In the binary mode, the 64 SiPM outputs are handled individually, first with a pre-amplifier and a fast-shaper. To simulate these two steps, we implemented electric models based on simplified circuits~\cite{marano}: the pre-amplifier is modeled as a 10\,MHz active low-pass filter, with an amplification factor of 10x, and the fast-shaper as a practical differentiator with a characteristic time of 15\,ns and a maximum gain of about 18x. The circuit components were selected to match the mentioned amplification and timing features and then tuned to obtain the best representation of the laboratory data used to validate the simulations (see section~\ref{sec:validation}).

The fast-shaper output is then processed with a discriminator and sampled with an FPGA into a 2048-bit trace. In a rough approximation, the discriminator outputs a signal of 3.3\,V if the fast-shaper is above the discriminator threshold and 0\,V otherwise. Then, the FPGA outputs a ``1'' if the discriminator signal is above 1.7\,V and a ``0'' if it is below 0.8\,V. In more detail, the discriminator has a typical rise time that affects the signal sampling: if one of the FPGA flip-flops is clocked when the discriminator signal is between 0.8\,V and 1.7\,V,  then the FPGA output is undetermined. To properly reproduce this in the simulation, we first estimated the rise time of the discriminator as the time that signal needs to rise between the baseline and 1.7\,V, using two thousand analog single-muon pulses measured at the output of the discriminator. We obtain a mean rise time of (1.51 $\pm$ 0.01)\,ns, and we assumed that if the FPGA samples during this rise time, then a ``0'' is output. On the other hand, if the fast-shaper output is above the discriminator threshold during more than 1.51\,ns when sampling, then a ``1'' is output. The transition time between 0.8\,V and 1.7\,V is rather negligible: different approaches were tested for the FPGA indetermination resulting all in the same outcome.

\begin{figure} [t]
	\centering
	\begin{subfigure}{1.0\textwidth}
		\centering
		\includegraphics[trim={0.0cm 0.0cm 0.0cm 0cm},clip,width=1.0\linewidth]{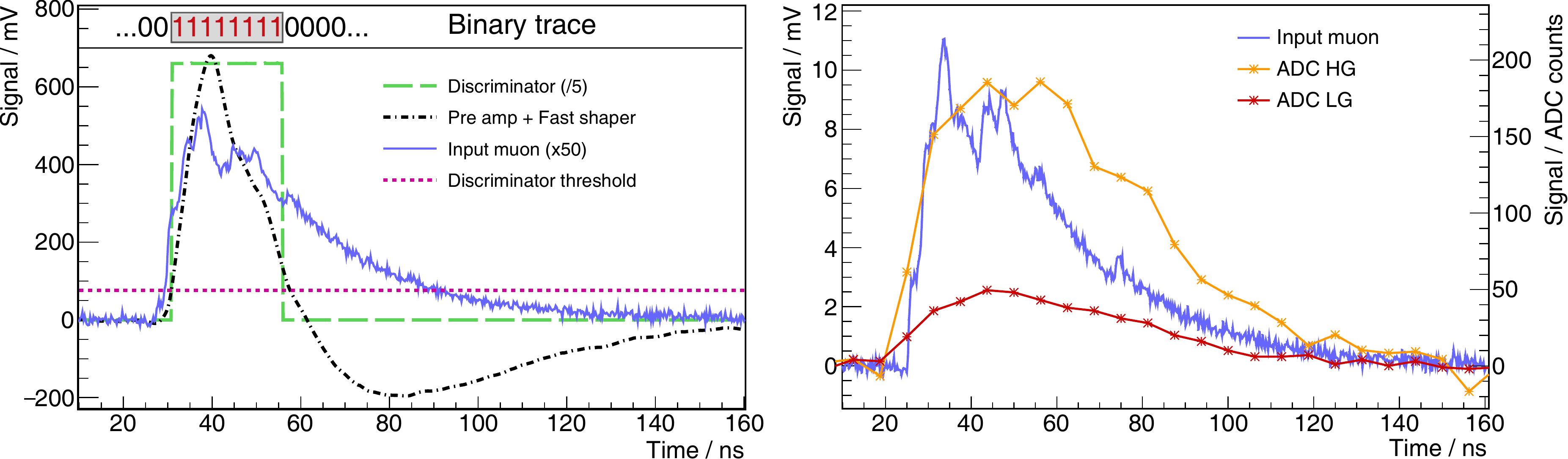}
	\end{subfigure}
	\caption{Two simulated single-muon signal at 2\,m for the binary (left) and ADC (right) modes. In the first case, the signal from the SiPM and the discriminator pulse are re-scaled for illustration.}
	\label{fig:exampleADC}
\end{figure}

In the left panel of Fig.~\ref{fig:exampleADC} we present an example of a simulated single-muon signal in the binary channel, where the re-scaled SiPM output (solid blue line) along with the output of the fast shaper (dotted-dashed black line) is displayed. 
The discriminator threshold set at 2.5\,PEs, as in the UMD modules, is presented (dotted pink line) along with the discriminator output (dashed green line) which was re-scaled to fit in the figure. When the fast-shaper output is above (below) the threshold during more than 1.51\,ns, at the output of the discriminator a 3\,(0)\,V signal is obtained and a ``1'' (``0'') is sampled in the binary trace for the corresponding time.

In the ADC mode, the electronics chain starts with two steps of adders that sum all the 64 SiPM analog signals, and an amplification step with a low- and high-gain amplifier follows. The amplifier outputs are sampled with an ADC at a speed of 160\,MHz, with a step of about 0.122\,mV/bit into two waveforms of 1024 samples. If we consider the components cutoff frequency to be the frequency for which the output of the circuit is $-3$\,dB (30\% fall in magnitude) of the nominal passband value, the first step of adders determines the signal timing: they introduce a frequency cutoff of about 12\,MHz~\cite{UMDFrontEnd}, while the other steps of the electronics introduce a frequency cutoff above 100\,MHz, with the bandwidth of the PE signal mostly between 15\,MHz and 2\,GHz. In this sense, the electronics response after the first step of adders is rather flat in frequency, and the whole transfer chain can be simplified to an active low-pass filter with a frequency cutoff of 12\,MHz, and with an amplification equivalent to the net gain of all the steps, $-2.50$($-8.47$)\,dB for the high(low)-gain channel~\cite{UMDFrontEnd}. Similar to the binary mode, these parameters were all tuned to match the laboratory data, and the result is presented in the next section.

An important feature in the ADC channel is the baseline fluctuation due to the SiPM dark counts, which has a significant impact on the signal charge estimation and detector resolution. For this reason, we included these fluctuations in the same way as we did for the PE analog signal: we determined the baseline offset and its fluctuations after the analog electronics processing and we input this information into a random generator that injects the fluctuations in the signal following a normal distribution. The baseline mean and standard deviation resulted in (8210.24 $\pm$ 0.02)\,ADC\,counts, and (3.24 $\pm$ 0.02)\,ADC\,counts for the low-gain channel, and in (8242.6 $\pm$ 0.1)\,ADC\,counts and (12.4 $\pm$ 0.1)\,ADC\,counts for the high gain.

The final step of the simulation of the ADC mode is the sampling, for which we grouped the amplifier outputs in bins of 6.25\,ns, and we extract the maximum amplitude reached in each bin. We then obtain the ADC\,counts per sample by diving the signal amplitude by the ADC step. An example of a simulated signal in the ADC channel is presented in the right panel of Fig.~\ref{fig:exampleADC}, where the SiPM output for one impinging muon produced at 2\,m is displayed (blue line and left y-axis), along with the output for the low- (orange) and high-gain (red) channels after the ADC sampling (stars and right y-axis).

\section{End-to-end validation of simulated data}
\label{sec:validation}

\begin{figure}[t]
	\centering
	\begin{subfigure}{1.00\textwidth}
		\centering
		\includegraphics[trim={0.0cm 0.0cm 0.0cm 0.0cm},clip,width=1.0\linewidth]{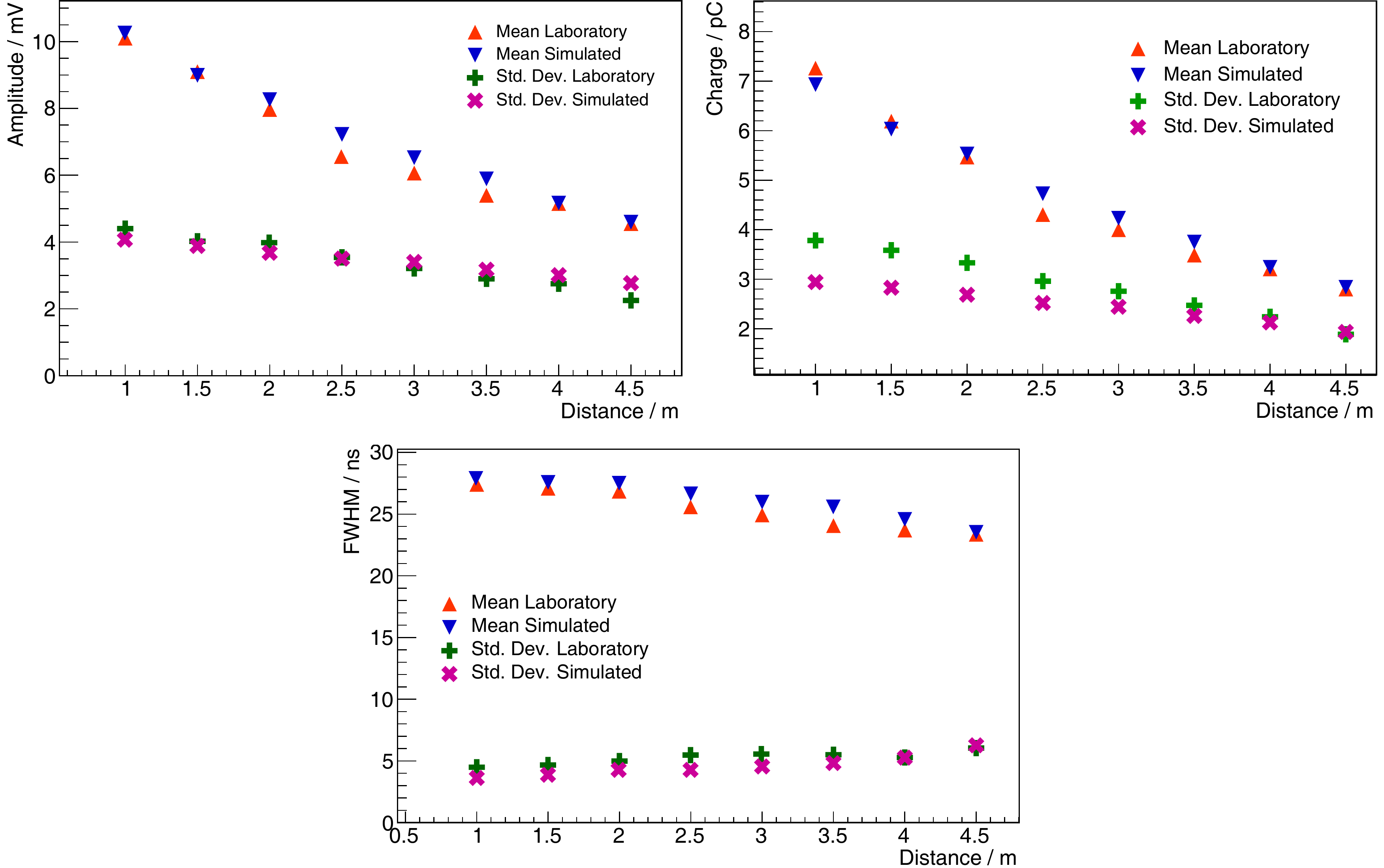}
	\end{subfigure}
	\caption{Main features of the muon signals at different positions on the scintillator strip. The means and standard deviation for laboratory data and simulations are shown. We display on the top-left panel the muon signal amplitude, in the top-right panel the muon signal charge, and in the bottom, the muon signal full width at half maximum (FWHM). The agreement between the data and the simulations is apparent.}
	\label{fig:muonPulsesSim}
\end{figure}

The UMD simulation must provide an accurate representation of the signal parameters relevant to the estimation of muon densities in air showers. In this section, we discuss the validation performed for both analog and digital signals to guarantee that this requirement is fulfilled. Considering the analog muon signals, the main features to assess are the maximum amplitude which has an impact on the performance of the binary mode, the signal charge, which has an impact on the ADC mode, and the signal width, which has an impact on the general timing of the signal. To verify that the analog simulation achieved in the previous section accurately describes these signal features, we present in Fig.~\ref{fig:muonPulsesSim} the maximum amplitude (top left), charge (top right), and full width at half maximum (bottom) of the single-muon signals as a function of the position where the muon impinged on the scintillator strip for two thousand muon signals at each position obtained with the muon telescope. In each panel, we present the results for the mean (orange up triangles) and standard deviation (pink crosses) measured in the laboratory, along with the mean (blue down triangles) and standard deviation (green plus signs) from the simulation. The average amplitude and charge of about 8\,mV and 5\,pC corresponds to the addition of the amplitude and charge of the PEs previously presented in Fig.~\ref{fig:average}: the amplitude (charge) of the muon signal is about 10\,(22)\,PE, each of which has an amplitude of 0.80\,mV, and a charge of 0.23\,pC. Similarly, the attenuation in charge and amplitude corresponds to the attenuation in the number of PEs presented in Fig.~\ref{fig:average}. The signal width is the result of the convolution between the time profile of the arriving photons and the time response of the SiPM.

Understanding the signal fluctuations is of utmost importance to assess the performance of the UMD. The variations in charge determine the resolution of the ADC channel to the number of impinging particles, while those in amplitude impact in the binary-mode efficiency, since signals that fluctuate to amplitudes below the discriminator threshold, do not produce ``1''s in the binary trace, and therefore, are not detected. In turn, the fluctuations in the signal width play a critical role when defining the reconstruction and calibration strategy for the UMD~\cite{UMDCalibration}. Therefore, the simulation needs to provide an accurate representation not only of the means of the parameters but also of their variation portrayed with the standard deviations of the distributions. From this stage on, (almost) no further fluctuations are introduced, since most stochastic processes involved in the UMD signal generation are found in the scintillator photo-production, the absorption and re-emission of the optical fiber, and in the SiPM photo-detection. The agreement between laboratory data and simulations is apparent in Fig.~\ref{fig:muonPulsesSim}, which denotes that the muon generator output provides a good description of the main features of the analog muon signals. The main differences between data and simulations are found in the charge fluctuation at distances close to the SiPM. These are produced by a few signals with a very low or large charge that are not reproduced in the simulation, which represents roughly 5\% of the data. Nonetheless, the fluctuations at this stage are compared to assess the general performance of the analog outputs and do not include the fluctuation in the baselines introduced by the SiPM thermal noise, which has a large impact at the end of the simulation chain. These are included at the last step of the simulation to recover the charge fluctuations relevant to the detector performance. The analog output is only the input of the read-out electronics, and as we showed in Fig.~\ref{fig:exampleADC}, after the processing, the traces become significantly different.

\begin{figure}[t]
	\centering
	\begin{subfigure}{0.45\textwidth}
		\centering
		\includegraphics[trim={0.0cm 0.0cm 0.0cm 0.0cm},clip,width=1.0\linewidth]{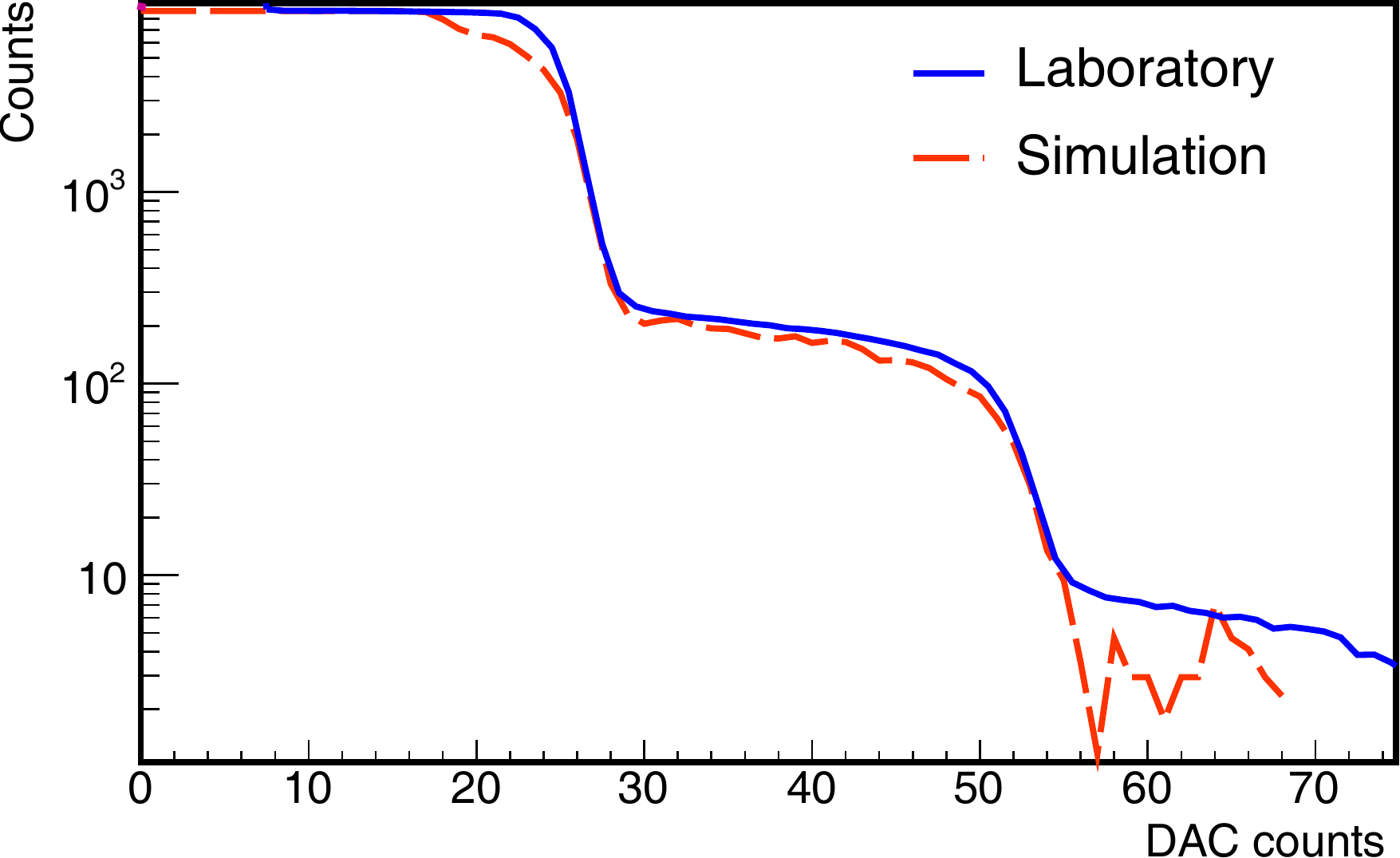}
	\end{subfigure}
	\caption{Dark-count rate curve used to calibrate the SiPM. We show the curves obtained using both laboratory data and the binary mode simulation. The single-PE amplitude obtained with the simulation is ($24.3 \pm 0.3$)\,DAC\,counts.}
	\label{fig:counterSignal}
\end{figure}

By implementing the transfer function of the binary mode, we can simulate dark-count rates as in Fig.~\ref{fig:calibrationPEGen}. Using the PE generator, we create pulses with a 2.2\% probability of having a cross-talk avalanche, and we processed the output with the simulation chain of the binary mode. To build the curve, we increase the count of the rate if the resulting binary trace has at least one ``1''. We compare, in Fig.~\ref{fig:counterSignal}, the simulated dark-count curve with the curve measured in the laboratory, for which we normalized the number of counts to the number of pulses generated with the simulation. As a result, we demonstrated that the procedure to obtain the PE amplitude can be accurately simulated obtaining similar values as in the laboratory; this procedure is the ground of the SiPM calibration. To obtain a ``1'' at the FPGA output the fast-shaper signal needs to be over the discriminator threshold during more than 1.51\,ns (see section~\ref{subsec:simeKIT}), due to the effect of the discriminator rise time. To better understand the impact of this effect on the SiPM calibration, we compared the single-PE amplitude obtained in the simulated calibration with that of the fast-shaper analog output. The mean PE amplitude is in this case (35.39 $\pm$ 0.04)\,mV, and for the simulated calibration (24.3 $\pm$ 0.3)\,DAC\,counts. Since the DAC step of the discriminator threshold equals 1.3\,mV, the amplitude obtained is (31.6 $\pm$ 0.4)\,mV. When the fast-shaper output barely reaches the discriminator threshold, the discriminator signal does not have time to rise and no ``1'' is recorded in the binary trace. It is then necessary to lower the threshold below the amplitude of the PE signal to effectively obtain a ``1'' in the binary trace; this dynamic is the reason for the underestimation of the PE amplitude with a negative bias of approximately 10\% when using the full chain of the binary channel. This 10\% bias does not represent a problem for the UMD calibration or performance, since the whole process for determining the optimal operation setup of the modules was performed within this same system~\cite{AMIGASIPM}, so the bias was consequently taken into account during the performance and calibration studies.

\begin{figure}[t]
	\centering
	\begin{subfigure}{1.00\textwidth}
		\centering
		\includegraphics[trim={0.0cm 0.0cm 0.0cm 0.0cm},clip,width=1.0\linewidth]{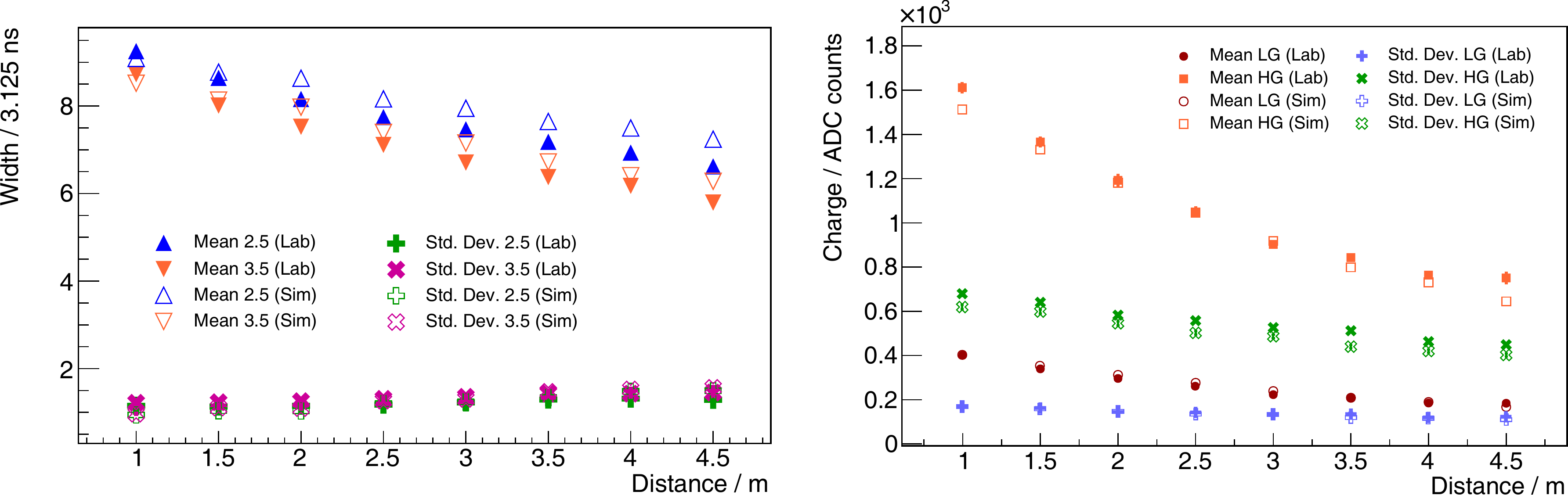}
	\end{subfigure}
	\caption{(Left) mean signal width over sample time and standard deviation in the binary mode as a function of the position on the scintillator strip with both laboratory and simulated data of single muons. The signal widths were obtained also using two different discriminator thresholds: 2.5 and 3.5\,PEs. (Right) Signal charge as a function of the position on the scintillator strip. We show the signal mean charge and standard deviation obtained using laboratory data. We also display the result using the ADC mode simulation.}
	\label{fig:counterResults}
\end{figure}

To obtain the number of muons using the binary channel, a reconstruction strategy based on the number of ``1'' in the traces needs to be determined to both reject the noise and preserve the detection efficiency as high as possible. To characterize the signal width, we present in the left panel of Fig.~\ref{fig:counterResults} the number of ``1''s as a function of the position on the scintillator strip obtained with the muon telescope. We present the signal mean widths (blue up triangles) and standard deviations (green plus signs), for laboratory measurements (full markers) and simulation (empty markers). Due to the optical-fiber attenuation, the signal mean width decreases when moving the muon telescope towards the end of the strips, which increases the total fluctuation over the whole strip and challenges the selection of the reconstruction strategy. If we consider all the widths within a three\,standard\,deviations from the mean, most of the muon signals in the laboratory have between 12 (37.5\,ns) and 4 (12.5\,ns) ``1''s. Furthermore, pulses produced by SiPM dark counts that reach the 2.5\,PE threshold have a typical width of less than 12.5\,ns~\cite{UMDCalibration,UMDICRC}. According to this, requesting a minimum width of 12.5\,ns effectively rejects 99\% of the SiPM noise without a significant loss of muon signals (about 1\%). With this reconstruction strategy, the probability of over-counting one muon in one event is reduced by a factor of about 2.7 with respect to a strategy in which a minimum width of 3.125\,ns (one sample) is requested to count as a muon. It is worth mentioning, that this estimations are valid in the laboratory where the effects of the soil or of clipping-corner muons are not taking place~\cite{MuonsWithAMIGA}.

\vspace{0.5cm}

The results shown in the left panel of Fig.~\ref{fig:counterResults} demonstrate that the simulation is accurately reproducing the binary traces measured in the laboratory: we obtained a maximum difference of less than 10\% in the mean width at the end of the strip, having the difference in the mean over the whole strip less than 4\%. To further validate the simulation, we have also performed this measurement with a discriminator threshold of 3.5\,PEs, for which the signal width is reduced on the order of about half a sample with respect to the 2.5\,PE threshold. We then present, also in the left panel of Fig.~\ref{fig:counterResults}, the results for the mean widths (orange down triangle) and standard deviation (pink crosses) of the laboratory data (full markers) and simulations (empty markers). The consistency between simulation and data denotes the potential to assess different configurations that eventually may be considered in the data taking for dedicated studies on the detector performance. It is also worth mentioning that when raising the discriminator threshold to 3.5\,PE, the number of SiPM dark counts reaching the threshold is reduced to 2.2\% with respect to the previous setup. In this sense, a muon can be considered as any ``1'' in the trace without significantly increasing the probability of over-counting a muon because of noise, but only equal to the previous result when considering three ``1''s.

The main feature in the ADC mode relevant to the UMD data reconstruction is the signal charge and its fluctuations presented in the right panel of Fig.~\ref{fig:counterResults}: to estimate the number of muons with the ADC channel we compute the signal charge and divide it by the mean charge of a single-muon. The mean signal charge for the low- (red circles) and high-gain (orange squares) channels along with their standard deviations (green crosses and blue plus signs, respectively) are presented as a function on the position of the scintillator strip for laboratory data (full markers) and simulations (empty markers). The signal charge was computed by adding the ADC\,counts in the muon signal of every sample above one standard deviation off of the baseline, and the average is presented for two thousand signals obtained using the muon telescope at each position. It is apparent how the PE attenuation has an important impact on the detector resolution since the total charge of single-muon signals between the beginning and end of the strip differs by a factor of two. The average charge over the whole strip for the low- and high-gain channels is (262 $\pm$ 5)\,ADC\,counts and (1059 $\pm$ 19)\,ADC\,counts with standard deviations of (141 $\pm$ 2)\,ADC\,counts, and (551 $\pm$ 8)\,ADC\,counts respectively. This translates into a resolution of about 60\% for single-muon signals in the ADC channel, which significantly diminishes with the arrival of many muons in the same module, the regime for which the ADC mode was designed.

Similar to the binary mode, the simulation of the ADC channel, presented in the right panel of Fig.~\ref{fig:counterResults}, accurately reproduces the signal charge and its fluctuations for both high- and low-gain channels. We obtained a maximum difference in the single-muon charge of less than 15\% in the high-gain channel, and less than 10\% in the low-gain channel for the mean at the end of the strip, being the difference in the average over the whole strip of less than 3\% for both means and standard deviations.

\begin{figure} [t]
	\centering
	\begin{subfigure}{1.00\textwidth}
		\centering
		\includegraphics[trim={0.0cm 0.0cm 0.0cm 0.0cm},clip,width=1.0\linewidth]{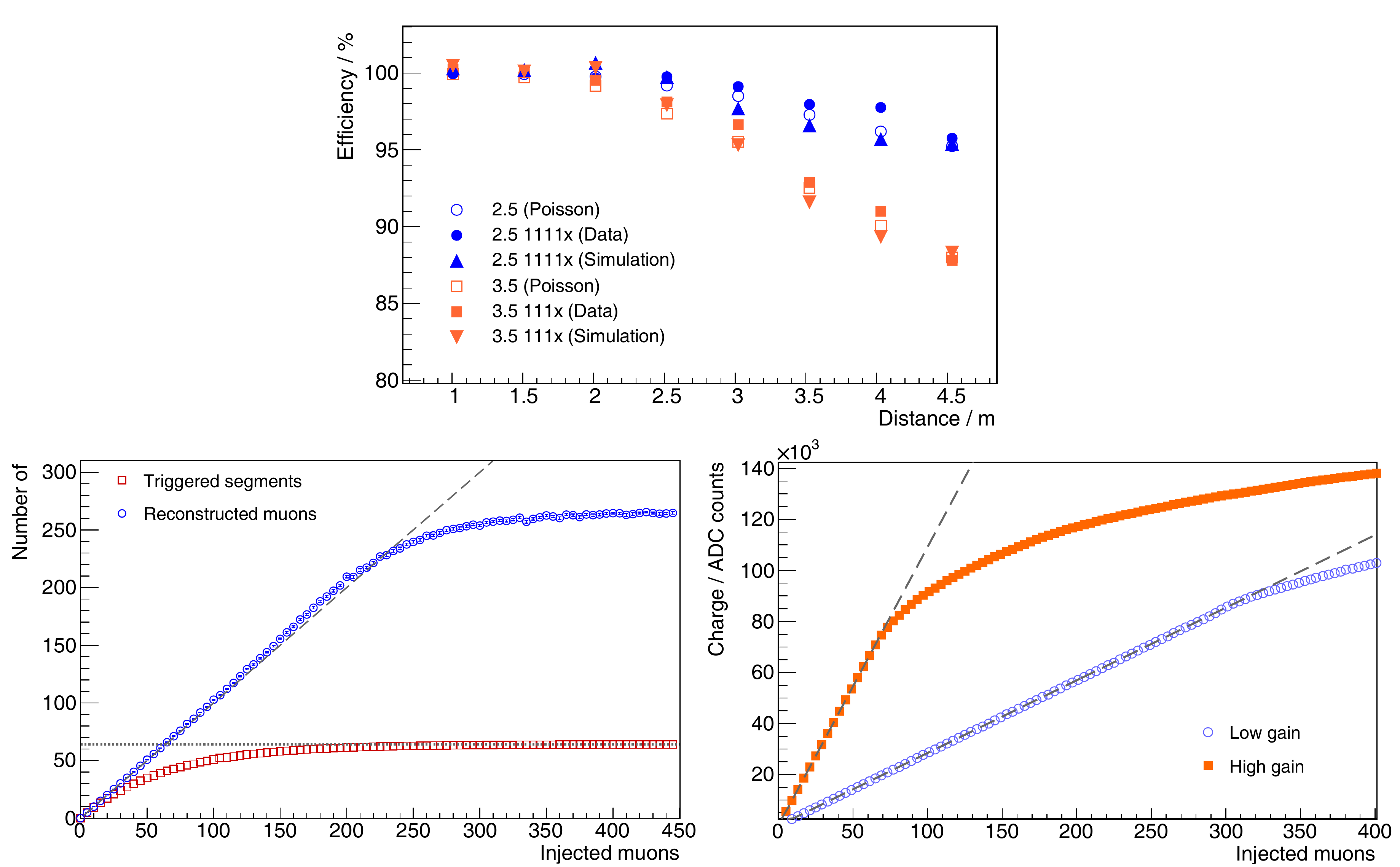}
	\end{subfigure}
	\caption{(Top) Efficiency of the binary mode as a function of the position on the scintillator strip estimated with laboratory data, simulations, and a Poissonian prediction using discriminator thresholds of 2.5 and 3.5\,PEs. (Bottom-left) number of triggered segments in the binary mode (red squares) and reconstructed muons (blue circles) after a pile-up correction in the binary mode as a function of the number of injected muons. The identity function is represented with a dashed-gray line. (Bottom-right) Mean signal charge of the low- (blue circles) and high-gain (orange squares) channels of the ADC mode simulation. To ease the interpretation we present in dashed lines the theoretical linear relation between the number of injected muons and signal charge.}
	\label{fig:efficiency}
\end{figure}

To reproduce the UMD performance realistically, the simulation must replicate the muon detection efficiency of the binary mode and the saturation of the ADC mode, which determines the dynamic range of the detector. In the top-left panel of Fig.~\ref{fig:efficiency} the detection efficiency of the binary mode for single muons is shown as a function of the position on the scintillator strip of the impinging particle. The efficiency is defined as the number of detected muons, using the reconstruction strategy previously described, over the total number of triggered events. It was estimated using laboratory data (blue full circles and orange full squares), simulations (blue up triangles and orange down triangles), and a Poissonian prediction (blue empty circles and orange empty squares) based on the measured number of PEs as a function of the position on the strip. This prediction consists of integrating a Poissonian distribution between the discriminator threshold (2.5 or 3.5\,PEs) and infinity, where the mean number of PEs at each position on the strip is extracted accordingly to measurements shown in Fig.~\ref{fig:average}.
The Poissonian prediction shows that the efficiency loss, mainly at the back of the strip, is principally produced by muon signals with amplitude fluctuating below the discriminator threshold. Therefore, the agreement between data and prediction denotes that there is not a significant efficiency loss in the signal processing or in the reconstruction analysis, having an integrated efficiency of 98.5\% over the whole scintillator strip\footnote{The discriminator threshold for the UMD is set at 2.5\,PE. Measurements with 3.5\,PE were obtained in the laboratory to validate the simulation with a different operating point, but are not foreseen in the field. The integrated detection efficiency, in this case, is roughly 96\%.}. In addition, it is worth mentioning that the simulation provides an accurate description of the binary mode efficiency for single muons.

To assess the saturation in the binary mode, we performed a toy Monte Carlo which consisted of injecting simultaneous particles randomly in a 64-segment detector and counting the number of segments hit by particles. We then determined the number of reconstructed particles by applying a pile-up correction~\cite{AMIGALDF} which acknowledges the probability of having multiple particles hitting the same segment. We present the results in the bottom-right panel of Fig.~\ref{fig:efficiency}: the mean number of triggered segments (red squares) and reconstructed muons (blue circles) as a function of the number of injected particles is shown, along with the identity function (dashed-gray line). The maximum number of segments that can be hit (dotted-gray line) corresponds to the 64 scintillator strips in a UMD module. At about 250 injected particles, the mean number of reconstructed muons begins to differ 5\% from the identity. It is worth mentioning that this toy Monte Carlo does not acknowledge for corner-clipping muons and knock-on electrons~\cite{MuonsWithAMIGA} from the soil that may produce additional hits on the segments.

Finally, in the bottom-right panel of Fig.~\ref{fig:efficiency} we show the mean signal charge of the low- (blue circles) and high-gain (orange squares) channels as a function of the number of muons injected simultaneously and simulated with the ADC mode. To ease the interpretation we present in dashed lines the theoretical linear relation between the number of injected muons and signal charge. To perform the simulation, the position of the muon on the scintillator strip was randomly drawn from a uniform distribution between 1\,m, and 5\,m. If we identified the saturation point where the signal charge is 5\% below the linearity with respect to the number of injected muons, the saturation for the high-\,(low-)gain channel occurs, roughly, at 81\,(357)\,muons per module. These results are quite consistent with previous tests in the laboratory~\cite{UMDFrontEnd} (85 and 362 respectively) where a light source was used to emulate the signal of several muons arriving at exactly the same time.

\section{Summary} 
\label{sec:conclusion}

\begin{table}[t]
\centering
\renewcommand{\arraystretch}{1.25}
 \begin{tabular}{c c c c c}
\toprule[1.25pt]
 \multicolumn{2}{c}{Parameter} & Laboratory & Simulation & Ratio\\
\midrule
 \multirow{2}{*}{Binary Width / 3.125 ns} & Mean & 7.75 $\pm$ 0.01 & 8.12 $\pm$ 0.03 & 0.95 \\
 & Std. Dev. & 1.49 $\pm$ 0.01 & 1.51 $\pm$ 0.02 & 0.99\\
 \multirow{2}{*}{Charge LG / ADC counts} & Mean & 262 $\pm$ 5 & 269 $\pm$ 3 & 0.97 \\
  & Std. Dev. & 141 $\pm$ 2 & 136 $\pm$ 2 & 1.04 \\
  \multicolumn{2}{c}{Binary efficiency } & 99\% & 98\% & 1.01\\
 \multicolumn{2}{c}{Saturation LG (in 10 m$^{2}$) / muons} & 362 & 357 & 1.01 \\
\bottomrule
 \end{tabular}
 \caption{Comparison of features relevant to the UMD event reconstruction and performance. The results are displayed for laboratory measurements and simulation, along with the ratio between them. }
 \label{table:summary}
\end{table}

Currently, the Pierre Auger Observatory is undergoing a major upgrade whose aim, among others, is to determine the composition of primary cosmic rays at the highest energies. The underground muon detector (UMD) is an essential part of this upgrade and is meant to obtain direct measurements of the muonic component in air showers for cosmic rays with energies between $10^{16.5}$ up to the region of the ankle (about $10^{18.7}$\,eV). To determine the number of hadrons in the primary particle using the information on the air shower muons, the UMD data ought to be properly interpreted at high-level physics analysis. To this aim, we needed to develop a reliable simulation tool that accurately reproduces the response of the detector to single particles.

In this work, we have presented a detailed description of the simulation sequence developed for the basic unit of the UMD detector: one scintillator strip with optical fiber and its corresponding optoelectronics. The models used in the simulation chain have been tuned and thoroughly validated using laboratory data. We present a summary of the main features relevant to the estimation of the number of impinging muons with the UMD and the detector performance in Table~\ref{table:summary}. The results for laboratory data and simulations, along with the ratio between these, are presented for the single-muon width and detection efficiency in the binary channel, and the signal charge and saturation in the ADC low-gain channel. The ratios show a maximum difference of 5\% proving the accuracy of the simulation at reproducing the relevant features of the signals. 

We have established that single-muon signals in the binary mode have a typical width between 12.5\,ns and 37.5\,ns. Using this as a condition to select muons, 99\% of the SiPM noise is effectively rejected while keeping an integrated efficiency of $99$\%. 
For the ADC mode, we have shown that the high-gain channel is capable of measuring up to about 80 muons arriving at exactly the same time, while the low-gain channel extends this number up to about 350 muons per 10\,m$^2$, results consistent with previous studies. Finally, it is worth mentioning that all the algorithms and models discussed in this article have been already implemented in \Offline, the official framework for analysis and simulation of the Pierre Auger Observatory~\cite{offline}.

\acknowledgments
The authors thank the support of the Pierre Auger Collaboration, in particular its Publication Committee for reviewing this article. They also thank the technical staff of the Instituto en Tecnologías de Detección y Astropartículas for aiding on the built of the setup used in this work. The corresponding author of this article has a post-doctoral grant from the Consejo Nacional de Investigaciones Científicas y Técnicas (CONICET) of Argentina.

\bibliographystyle{JHEP}
\bibliography{UMDSim}{}


\end{document}